\documentclass{PoS}
\usepackage{enumitem}
\title{Gamma--Ray Burst observations with Fermi}
\ShortTitle{GRB observations with Fermi}
\author{
\speaker{E.~Bissaldi}$^{\;(1)\;}$, 
F.~Longo$^{\;(2,3)\;}$, 
N.~Omodei$^{\;(4)\;}$, 
G.~Vianello$^{\;(4)\;}$,
A.~von Kienlin$^{\;(5)\;}$
 $\quad$ $\quad$ $\quad$ $\quad$ $\quad$  $\quad$ $\quad$ $\quad$ $\quad$ $\quad$
$\;\;\;\;$on behalf of the Fermi/LAT Collaboration \\
$^{(1)}$ Istituto Nazionale di Fisica Nucleare, Sezione di Bari, 
Bari, Italy \\
$^{(2)}$ Dipartimento di Fisica, Universit\`a degli Studi di Trieste, 
Trieste, Italy \\
$^{(3)}$ Istituto Nazionale di Fisica Nucleare, Sezione di Trieste-Udine,
Trieste, Italy \\
$^{(4)}$ Stanford University, Stanford, CA, United States \\
$^{(5)}$ Max-Planck-Institut f\"ur extraterrestrische Physik, Garching, Germany \\
{\footnotesize{E-mail:}} {\tt{\footnotesize{Elisabetta.Bissaldi@ba.infn.it}}}}
\abstract{
After seven years of science operation, the 
{\it Fermi} mission has brought great advances in 
the study of Gamma-ray Bursts (GRBs). Over 1600 GRBs 
have been detected by the Gamma-ray Burst 
Monitor, and more than 100 of these are 
also detected by the Large Area 
Telescope above 30 MeV.

We will give an overview of these observations, 
presenting the common properties in the GRB temporal 
and spectral behavior at high energies. 
We will also highlight the unique characteristics 
of some individual bursts. The main physical implications 
of these results will be discussed, along with open questions 
regarding GRB modeling in their prompt and 
temporally--extended emission phases.}
\FullConference{The 34th International Cosmic Ray Conference,\\
		30 July- 6 August, 2015\\
		The Hague, The Netherlands}
\begin{document}
\section{The \textit{Fermi} mission}
\label{Fermi}
The contribution of the \textit{Fermi} mission
to the study of the high--energy emission from GRBs
has been of great importance over the past seven years.
Thanks to its 12 NaI detectors and two BGO detectors, 
sensitive in the 8 keV--1 MeV and 250 keV--40 MeV energy ranges,
respectively, and to an almost 4$\pi$ sr Field of View (FoV), 
the Gamma--Ray Burst Monitor (GBM) \cite{MEE09} has
detected over 1600 bursts at the time of this writing.

A great number of GBM GRBs has also been detected at higher energies 
by the Large Area Telescope (LAT) \cite{ATW09}.
This pair production gamma--ray telescope is
sensitive from $\sim$ 30 MeV to $>$300 GeV and features 
a FoV of 2.4 sr at 1 GeV, a low dead time per event 
(27 $\mu$s) and the largest effective area for 
gamma--ray space satellites at GeV 
energies. The actual LAT performances can be found in 
\cite{LATperf}.
This allows the LAT to get a larger number 
of GRB detections at energies  $>\sim$30  MeV 
($\sim$9 GRBs/year) with respect to its predecessor 
EGRET (5 GRBs in 10 years) or to AGILE 
(7 GRBs in 8 years), and in rough agreement with the 
pre--launch expectations \cite{BAN09}.
\begin{figure}[b!]
\centering
\includegraphics[width=0.9\textwidth,bb=0 0 635 306]{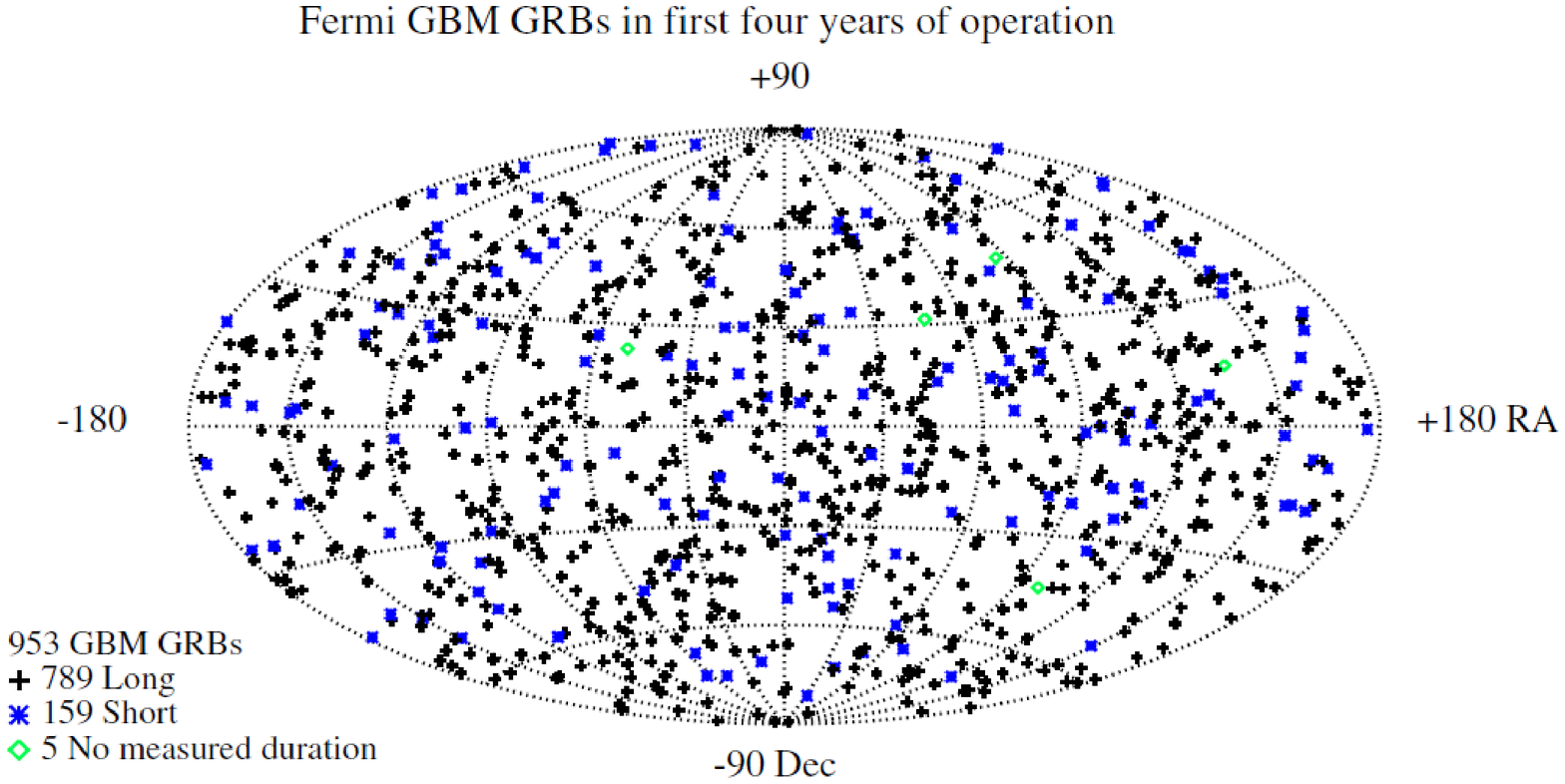}
\caption{Sky distribution of GBM triggered GRBs in celestial 
coordinates. Crosses indicate long GRBs (T90 > 2 s); 
asterisks indicate short GRBs. Taken from \cite{AVK14}.}
\label{fig_map_GBM}
\end{figure}
\section{The \textit{Fermi}/GBM GRB catalogs}
The most important observables of the GBM detected 
GRBs are summarized in a series of catalogs.  
The latest catalog issue, the second GRB 
catalog \cite{AVK14}, lists 953 GRBs
detected in the first four years of the mission, and 
provides their locations, the main characteristics 
of the prompt emission, the durations, peak fluxes and fluences. 
A sky distribution of these GRBs
is shown in Figure \ref{fig_map_GBM}.
An accompanying second spectral catalog \cite{GRU14} provides 
information on the systematic spectral analysis 
of nearly all GRBs listed in the trigger catalog. 
Time-integrated fluence and peak flux spectra are 
presented for all GRBs. Moreover,
a separate catalog reporting time resolved spectral 
analysis of the brightest GBM GRBs is in preparation.
Preliminary results of this type of analysis
on eight bright GRBs can be found in \cite{YU15}.
All of these catalogs are an official product 
of the {\it Fermi}/GBM science team, and the data 
files containing the complete results are 
available from the High-Energy Astrophysics 
Science Archive Research Center (HEASARC)
\cite{HEAS}.

The GRB sample analysis presented in the second catalog establishes 
the conclusions of the first catalog, which covered
two years of data acquisition \cite{PAC12}. 
The rate of burst detections ($\sim$240 GRBs/yr), 
which is only slightly smaller compared to the rate 
of the BATSE instrument ($\sim$300 GRBs/yr), can 
be explained by the GBM additional range of 
trigger timescales (primarily the 2 s and 
4 s timescales), which are compensating 
for the higher burst detection threshold of 
GBM ($\sim$0.7 versus $\sim$0.2 photons 
cm$^{-2}$ s$^{-1}$ for BATSE). The distribution 
of GBM durations is consistent with the 
well--known bimodality measured previously 
and the fraction of about 17\%\ of short 
GRBs in the GBM sample is somewhat 
smaller than detected by BATSE, 
which is mainly attributed to the GBM 
capability to trigger on longer timescales. 

The spectral properties of the bursts
presented in the spectral catalog are described
both from a time--integrated and a peak--flux analysis. 
Four photon model fits are applied to each selection, 
resulting in more than 7500 spectral fits. 
Model comparison techniques were applied
to select the most statistically preferred 
model for each GRB and 
the resulting GRB parameter distributions
were investigated extensively.
Small discrepancies between the
first GBM spectral catalog \cite{GOL12} 
and the second spectral catalog are mainly due 
to enhanced and modified criteria in 
determining the GRB samples, updated analysis 
software and response matrices and
the usage of different statistical criteria.
\section{The \textit{Fermi}/LAT GRB catalog}
The first \textit{Fermi}/LAT GRB catalog
\cite{ACK13} covers three years of 
observations, from August 2008 to July 2011. 
In this period GBM detected 
$\sim$750 GRBs, with around half of them 
in the LAT FoV. The two detection algorithms used
by LAT are: (1) a standard likelihood algorithm, 
providing both detection and localization 
with $< 1$ deg accuracy, using the post--launch 
reconstruction method
{\it ``Pass 6 v3 Transient''} \cite{RAN09} 
for events above 100 MeV; and (2) a counting 
analysis using the LAT Low--Energy data (LLE) 
\cite{PEL10}. 
The LLE technique features a large effective 
area starting at $\sim$30 MeV but no localization 
capability. 
While algorithm (1) detected and localized 28 GRBs, algorithm
(2) uncovered 7 more bursts, for a total of 35 GRBs.

While the number of GRBs detected
by LAT is a small fraction of the total number 
of GBM GRBs in the LAT FoV, this sample allows 
us to uncover unique features of GRBs 
emerging only at high energies.
\subsection{Energetics}
Since LAT observations are photon--limited rather 
than background--limited, the detection efficiency 
is directly related to the counts fluence 
of the source. This is an important difference 
with respect to GBM, which is background limited 
and for which the peak flux of the source 
is more relevant. Of course, the low--energy 
fluence is highly correlated with the high--energy fluence. 
The fact that LAT detects preferentially GRBs
with a high low--energy fluence is therefore 
not surprising. The typical ratio between the 
high--energy fluence (above 100 MeV) and the 
low--energy fluence (10 keV--1 MeV) is $\sim 0.1$. 

It is interesting to note the catalog 
presents four \textit{hyper--energetic} bursts for 
which the ratio greatly exceeds the typical value, being 
closer or even above 1. 
The same conclusion can be reached taking the ratio 
of the rest frame total energy $E_{iso}$ 
in the two energy bands, which demonstrates that 
this is not an effect of the distance of these bursts,
which are distributed between redshift 0.9 and 4.35. 
\subsection{Delayed and temporally extended emission}
The emission above 100 MeV is systematically delayed 
with respect to the low--energy emission.  
When using $T_{05}$ as a measure of the onset of 
the emission for both the 10--300 keV energy 
range (from \cite{PAC12}) and 
the 100 MeV--10 GeV energy range, 
it is clear that the latter is systematically 
larger than the former. Also, the duration of the
high--energy emission appears to be systematically 
longer, and features a smooth decaying phase 
after the end of the low--energy prompt emission. 

Such decaying phase is well described by a power 
law with typical decay index $\alpha_{L} \simeq -1$, pointing 
to a different physical origin with respect to
the spiky prompt emission. This value is foreseen 
by the standard afterglow model for an 
adiabatic expansion of the fireball. A
radiative expansion would foresee a decay with 
an index of 10/7, which is not observed in the LAT data.

In a few cases, a broken power law gives a 
better description of the temporally extended data,
and the time of the temporal decay break is usually
found after the end of the low--energy emission, 
as measured by T$_{90}$. 
\subsection{High-energy photons}
The LAT has observed photons up to 95 GeV 
coming from bright GRBs, which in the case of 
high--redshift GRBs can become more than 
100 GeV in the rest frame of the progenitor 
of the burst. This result poses a big 
challenge for the efficiency of the 
particle acceleration mechanisms, 
especially when considering the fact 
that some of these high--energy events 
have been detected within seconds since 
the start of the low--energy emission. 
In the context of the standard fireball 
model \cite{PIR99},
the presence of such high--energy photons 
constrains also the bulk Lorentz 
factor of the emitting shells to 
be $\Gamma > 1000$ in some cases, 
a value much higher than what previously thought. 

High--energy photons from high--redshift GRBs also
allow to constrain the opacity 
of the Universe connected with the interaction 
of the $> 10$ GeV $\gamma$--rays with 
optical and UV photons of the Extragalactic 
Background Light (EBL). In the case of 
the short GRB 090510, the short time delay 
observed between low and high-energy 
events can be used to place tight 
limits on the energy dependence of 
the speed of light, which is postulated
 for example by some quantum gravity 
theories \cite{VAS13}.
\subsection{Broad--band spectroscopy}
\textit{Fermi} is an exceptional 
observatory for GRB spectroscopy of the prompt emission. 
By combining GBM and LAT data, we can achieve
an unprecedented spectral coverage
spanning almost 7 decades in energy,
from 10 keV up to 300 GeV. 
This feature was exploited in the catalog 
by performing a broad--band spectral 
analysis of all the GRBs contained in the sample.
Before \textit{Fermi} most of the GRB 
spectra were well described by the 
empirical Band model \cite{BAN93}, 
which has become the \textit{de--facto} 
``standard GRB model''. 
The spectra of all the brightest bursts 
inside the LAT FoV present, 
on the contrary, significant deviations 
from a Band function, 
requiring additional components such 
as power laws, high--energy cutoffs, 
or both. Other GRBs, observed at 
low off--axis angles, and with a corresponding 
high effective area, show deviations as well. 
We conclude that the empirical Band model 
seems to be not sufficient to describe all 
the spectral features of LAT GRBs. 
Unfortunately, there is no common 
recipe, and different components can be 
required depending on the particular event. 
This calls for a better broad--band 
modeling of the prompt spectra of GRBs, 
opening new questions and prompting new 
theoretical developments.
\subsection{The afterglow of LAT--detected GRBs}
A subsample of LAT--detected GRBs have been 
studied at other wavelengths, in particular 
during their afterglow emission. A systematic 
study published by \cite{RAC11} 
shows that in many ways the properties of 
the afterglow of LAT bursts are typical 
of the general afterglow population, but the 
ratio between the luminosity of the prompt 
emission and the luminosity of the afterglow 
is larger. Therefore, either their prompt 
emission is more efficient in producing 
gamma--rays, or, conversely, their afterglows 
are somehow suppressed. 
In two cases, GRB 090510 and GRB 110731A, 
{\it Swift} and other instruments observed 
the afterglow when the high--energy 
extended emission was still detectable 
by the LAT. Dedicated broadband studies, from 
optical wavelengths to gamma--rays, 
showed that the emission is compatible 
with being from external shocks
\cite{DEP10,ACK13b}. In another case, 
GRB 100728A, high--energy emission was 
detected by the LAT only in correspondence 
with an X--ray flare, which was successfully 
modeled from X to gamma--ray energies as internal 
shock emission \cite{ABD11}.
\section{The extremly bright GRB 130427A}
The observations of the exceptionally 
bright GRB 130427A by \textit{Fermi}
provide further constraints on the GRB 
phenomenon and their emission processes. 
It is one of the most energetic GRBs ever observed. 
The initial pulse up to 2.5 seconds is possibly 
the brightest well--isolated pulse observed to date. 
A fine time resolution spectral analysis led by the
{\it Fermi}/GBM Team \cite{PRE14} highlighted how
difficult it is for any of the
existing models to account for all of the observed 
spectral and temporal behaviors simultaneously.

Furthermore, GRB 130427A had the largest fluence, 
highest observed energy photon (95 GeV), 
longest gamma--ray duration (20 hours), 
and one of the largest isotropic energy
releases ever observed from a GRB. 
The temporal and spectral analyses 
of GRB 130427A presented by the {\it Fermi}/LAT 
Team in \cite{ACK14} challenge the widely 
accepted model that the non--thermal high--energy 
emission in the afterglow phase of GRBs 
is synchrotron emission radiated by electrons 
accelerated at an external shock.
\section{Towards the second LAT GRB catalog}
All LAT GRB detections are constantly 
maintained and kept up--to--date on two publicly 
available GRB tables \cite{table,asdc}. They list
100 GRBs at the time of this writing.
The \textit{Fermi}/LAT collaboration is actively 
working to produce the second version of its 
GRB catalog, probably covering six years into the mission. 
This catalog will contain more GRBs, 
not only due to an extended period of data acquisition, 
but also due to (1) renewed algorithms to search 
on a wider angular region centered on 
GBM GRB trigger positions and to (2) new LAT 
data selection, with larger effective area 
both at low and at high energies. 
\subsection{New GRB detection algorithm}
Starting from the results obtained in the first LAT
GRB catalog and thanks to recent developments
in the understanding of the systematic
errors on GBM localizations \cite{CON15},
we are currently developing and testing 
a new detection algorithm which 
increases the number of detections
by $>$45\%\ and was first presented in \cite{VIA15}.  
It consists of 10 searches running 
in parallel over time intervals logarithmically 
spaced from the trigger time to 10 ks after that. 
In case of a GBM trigger, for each of these time intervals
the algorithm creates a grid
in equatorial coordinates and with a spacing of
0.7 deg, covering a finding map
of 30$\times$30 deg, thus covering the GBM
position uncertainty.
For each point of the grid, a likelihood analysis is
performed and a test statistics (TS) value is computed
Finally, the maximum of the TS in the map
is considered as the best guess for the position of
the new transient, and marked for further position
optimization. A new likelihood analysis is performed on the
best position found. If the TS from this final
analysis is above a certain threshold, we consider
it a new detection (a 5$\sigma$ detection corresponds to 
TS$\sim$28).
\begin{figure}[b!]
\centering
\includegraphics[width=0.8\textwidth,bb=0 0 601 442]{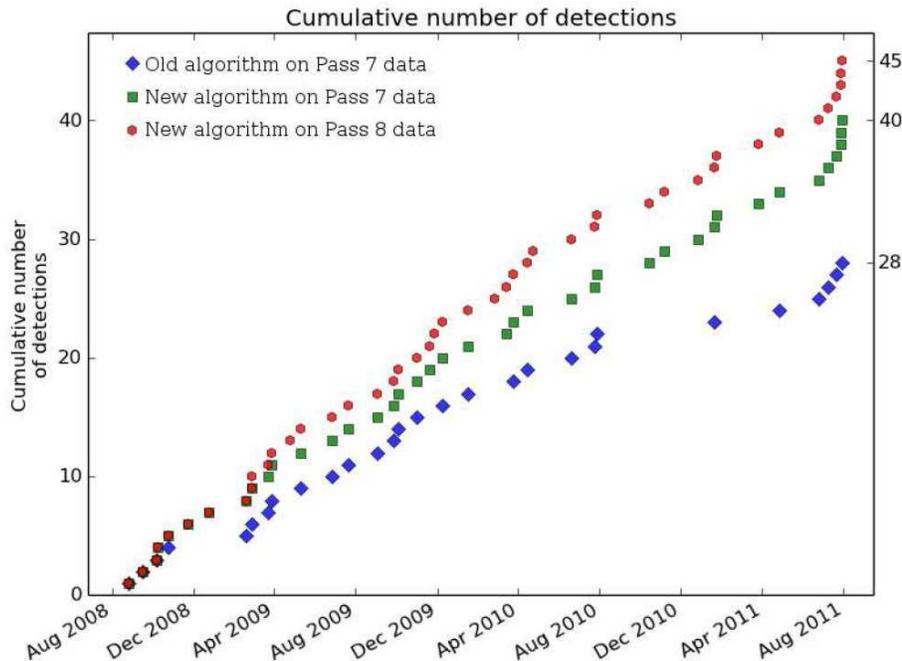}
\caption{Cumulative detections for the time span covered by the LAT GRB catalog. The new analysis yields 45\%\ more detections than the one used in the catalog when run on the same data ({\it blue} and {\it green}), and
60\%\ more with Pass 8 data ({\it red}). Taken from \cite{VIA15}.}
\label{fig_new_GRB}
\end{figure}
\subsection{GRBs with Pass 8}
Since 2010, the {\it Fermi}/LAT collaboration is 
developing a comprehensive revision of the event--level 
analysis, known as ``Pass 8''.
It provides a full reprocessing 
of the entire mission dataset, including 
improved event reconstruction, a wider energy 
range, better energy measurements, and 
significantly increased effective area. 
In addition, the events have been evaluated 
for their measurement quality in both position and energy. 
As of the release date (June 24, 2015), the Fermi
Sciene Support Center (FSSC) is now serving Pass 8 LAT data 
for analysis \cite{LATweb}. 

Using its preliminary achievements, the 
LAT collaboration re--analyzed the prompt phase 
of ten bright GRBs previously detected by 
the LAT \cite{ATW13}, finding four new gamma rays 
with energies greater than 10 GeV in 
addition to the seven previously known. 
Among these four there is a 
27.4 GeV gamma-ray from GRB 080916C, 
which, at a redshift of 4.35, makes 
it the gamma--ray with the highest intrinsic 
energy (147 GeV) detected so far 
from a GRB. 

If we combine the specialized search algorithm
with the Pass 8 event selection, we largely
enhance the detection 
of faint high--energy GRB counterparts, 
increasing the efficiency of detection by more
than 60\%, yielding more than 100
bursts over the time span of the {\it Fermi} mission
(see Figure \ref{fig_new_GRB}). 
The analysis and characterization of this new 
sample is currently in progress. 
When completed, it will hopefully help to settle
some open questions which we could not firmly 
establish in the first catalog due to the 
limited statistics, such as the existence 
of a separate population of hyper--energetic events,
the nature of the fireball expansion (adiabatic vs.~radiative), 
or how common are spectral cutoffs.
\acknowledgments
The \textit{Fermi}--LAT Collaboration acknowledges 
support for LAT development, operation and
data analysis from NASA and DOE (United States), 
CEA/Irfu and IN2P3/CNRS (France), ASI
and INFN (Italy), MEXT, KEK, and JAXA 
(Japan), and the K.A. Wallenberg Foundation, 
the Swedish Research Council and the National 
Space Board (Sweden). Science analysis support in
the operations phase from INAF (Italy) and 
CNES (France) is also gratefully acknowledged.
\end{document}